\documentstyle[preprint,revtex]{aps}

\begin{document}
\draft
\preprint{Alberta Thy--18--93}
\begin{title}
Chiral symmetry and many-body nonleptonic decays of heavy hadrons
\end{title}
\author{Q.\ P.\ Xu  and A.\ N.\ Kamal}
\begin{instit}
Theoretical Physics Institute and Physics Department\\
University of Alberta\\
Edmonton, Alberta, Canada, T6G 2J1
\end{instit}
\begin{abstract}
We suggest that, similar to the many applications of chiral
effective Lagrangian made recently to semileptonic decays of
heavy hadrons involving Goldstone bosons,
chiral symmetry can also be applied to many-body
nonleptonic decays of heavy hadrons.
As examples we establish the chiral effective Hamiltonian for some
typical many-body nonleptonic decays of bottom hadrons. We
discuss the lowest-order contributions coming from such a
Hamiltonian. We emphasize that wide applications of this
method are possible.
\end{abstract}
%\pacs{PACS numbers: 13.30.Eg, 14.20.Kp}
%\narrowtext
\newpage

\begin{center}
{\bf 1. Introduction}
\end{center}

Recently, there has been progress in establishing a chiral
perturbation theory incorporating heavy quark symmetry
to study the interactions
of heavy hadrons with Goldstone bosons \cite{r1,r2,r3}.
This method has been applied to various processes, in particular,
to semileptonic decays such as
$B \rightarrow \pi e \bar \nu$,
$B \rightarrow D\pi e \bar \nu$ and
$\Lambda_b \rightarrow \Lambda_c\pi e\bar \nu$ \cite{r1,r2,r3,r4}.
Though the Goldstone bosons in these decays are
quite energetic in most kinematic region, there is
always a region of the Dalitz-plot
where the Goldstone bosons are soft and where it is
possible to apply chiral symmetry.

In this letter, we point out the possibility of similar applications
of chiral effective Lagrangians to many-body nonleptonic decays of
heavy hadrons involving Goldstone bosons.
As far as we know, little attention has been paid to this type of
applications. In \cite{r4b}, two-body heavy-flavor-conserving
nonleptonic decays of charmed baryons has been studied using
the chiral effective Lagrangian. Here we look at the
heavy-flavor-changing nonleptonic decays of heavy hadrons.
In two-body nonleptonic decays of heavy hadrons, the
Goldstone boson momenta are fixed and are quite large. However, in
three(or more)-body nonleptonic decays, there is also always a
region where the Goldstone bosons involved have low
energies. Chiral symmetry can be applied in this region.
% An example is $B\rightarrow \pi e \bar \nu$.
Among the hadronic decays there are many such examples,
for instance, $B \rightarrow D D_s \pi$,
$\Lambda_b \rightarrow \Lambda_c D K$ etc.

\begin{center}
{\bf 2. Chiral effective Lagrangian for heavy hadrons}
\end{center}

The chiral effective Lagrangian for heavy hadrons
containing one heavy quark has been studied by
several authors \cite{r1,r2,r3,r5}.

The lowest-order chiral effective Lagrangian of heavy hadrons
(meson and baryon) which incorporate both heavy quark symmetry
and chiral symmetry is \cite{r1,r2,r3,r5}
\FL
\begin{eqnarray}
{\cal L}^{(0)}=&&-i Tr_D( {\bar H}_i v\cdot D H_i)+
g_1 Tr_D( {\bar H}_i H_j {\not\!\!A }_{ji}\gamma_5 )\nonumber\\
&&-i\ Tr_G( {\bar B^\mu_6}\ v\cdot D B_{6\mu})
-\frac{1}{2}i\ Tr_G( {\bar B_{\bar 3}}\ v\cdot D B_{\bar 3})
\nonumber\\
&&+i\ g_2 \epsilon_{\mu\nu\alpha\beta}
Tr_G ({\bar B^{T\mu}_6} v^\nu A^\alpha B^\beta_6)+
g_3 Tr_G ({\bar B_{\bar 3}} A_\mu B^\mu_6+
{\bar B^\mu_6} A_\mu B_{\bar 3} )
\label{eq:e1}
\end{eqnarray}
Higher-order terms which break the chiral symmetry and heavy
quark symmetry can be added systematically \cite{r1,r5}.
However, here we work only in the leading order in both
the Goldstone momentum and $1/m_Q$ expansions ($m_Q$ is the heavy
flavor mass). The notation in (\ref{eq:e1}) is the following:
For meson filed $H_i$ with heavy quark $Q$
($i$ is the light-quark index, the so-called `` Goldstone index "),
\FL
\begin{eqnarray}
(H_1, H_2, H_3)=(Q \bar u, Q \bar d, Q \bar s)
\label{eq:e2}
\end{eqnarray}
$H_i$ is also a $4\times 4$ matrix with respect to the Dirac
index (See, for example, \cite{r1})
\FL
\begin{eqnarray}
H_i=\frac{1+\not\!v}{2} (-P_i \gamma_5+P_{i \mu}^\ast \gamma^\mu)
\label{eq:e3}
\end{eqnarray}
where $P_i$ nad $P_{i \mu}^\ast$ are the pseudoscalar and vector
meson fields with velocity $v$.
For D-mesons, $(P_1, P_2, P_3)=(D^0, D^+, D^+_s)$ and
$(P_1^\ast, P_2^\ast, P_3^\ast)=
(D^{0^\ast}, D^{+^\ast}, D^{+^\ast}_s)$. For B-mesons,
$(P_1, P_2, P_3)=(B^-, {\bar B}^0, {\bar B}^0_s)$ and
$(P_1^\ast, P_2^\ast, P_3^\ast)=
(B^{-^\ast}, {\bar B}^{0^\ast}, {\bar B}^{0^\ast}_s)$.
$\bar H_i$ is defined as
\FL
\begin{eqnarray}
\bar H_i=\gamma_0 H^+_i \gamma_0, \ \
\bar H_i= (P^{+}_i \gamma_5+P_{i\mu}^{\ast +} \gamma^\mu)
\frac{1+\not\!v}{2}
\label{eq:e4}
\end{eqnarray}
$Tr_D$ implies a trace over Dirac indices and $Tr_G$ that over
the Goldstone indices.
The covariant derivative, $D^\mu$, is defined as
\FL
\begin{eqnarray}
D^\mu H_i=\partial^\mu H_i - H_jV^\mu_{ji}
\label{eq:e6}
\end{eqnarray}
with
\FL
\begin{eqnarray}
V^\mu=\frac{1}{2} (\xi^+ \partial^\mu \xi+\xi \partial \xi^+ )\ .
\label{eq:e7}
\end{eqnarray}
The axial vector field $A_\mu$ in (\ref{eq:e1}) is
\FL
\begin{eqnarray}
A^\mu=\frac{i}{2} (\xi^+ \partial^\mu \xi-\xi \partial \xi^+ )\ .
\label{eq:e8}
\end{eqnarray}
$\xi$ is the $3\times 3$ matrix for Goldstone bosons:
\FL
\begin{eqnarray}
\xi=\sqrt {\Sigma}, \ \ \ \ \Sigma=exp(\frac{2iM}{f}),\ \ \ M=
 \left( \begin{array}{ccc}
\frac{\textstyle \pi^0}{\textstyle \sqrt2}+
\frac{\textstyle \eta}{\textstyle \sqrt6}& \pi^+& K^+\\
\pi^-&-\frac{\textstyle \pi^0}{\textstyle \sqrt2}+
\frac{\textstyle \eta}{\textstyle \sqrt6}&K^0 \\
K^-& \bar K^0 & \frac{\textstyle -2}{\textstyle \sqrt6}\eta
    \end{array}  \right)\
\label{eq:e9}
\end{eqnarray}
with $f\simeq 0.132$GeV.

% The second term in (\ref{eq:e1}) describes the interaction of
% heavy meson $H_a$ with Goldstone boson $\xi$.

The third to the last term in (\ref{eq:e1}) represent interactions
associated with heavy baryons containing a single heavy quark.
Single heavy quark baryons can be classified by their light quark
content. The two light quarks can be either in a symmetric
$SU(3)$ sextet ${ 6}$ with spin 1 or in a antisymmetric triple
${\bar 3}$ with spin 0. Thus antitriplet baryon field $B_{\bar 3}$
has spin $1/2$ while sextet baryon field $B_6$ has spin either
$1/2$ or $3/2$. With respect to Goldstone index, $B_{\bar 3}$
and $B_6$ are $3 \times 3$ matrices. For charmed baryons,
for instance,
\FL
\begin{eqnarray}
B^{(c)}_{\bar 3}=
\left( \begin{array}{ccc}
0& \Lambda_c^+&\Xi^+_c\\
-\Lambda_c^+& 0 & \Xi^0_c \\
-\Xi^+_c & -\Xi^0_c& 0
 \end{array}  \right)\      \ \ \ \ \ \
B^{(c)}_6 ( {\textstyle spin} 1/2)=
\left( \begin{array}{ccc}
\Sigma^{++}\ \ &
\frac{\textstyle \Sigma^+_c}{\textstyle \sqrt2}\ \ &
\frac{\textstyle \Xi^{+\prime}_c}{\textstyle \sqrt2}\ \\
\frac{\textstyle \Sigma^+_c}{\textstyle \sqrt2} \ \ &
 \Sigma^0_c \ \ &
\frac{\textstyle \Xi^{0\prime}_c}{\textstyle \sqrt2}\ \\
\frac{\textstyle \Xi^{+\prime}_c}{\textstyle \sqrt2}\ \ &
\frac{\textstyle \Xi^{0\prime}_c}{\textstyle \sqrt2}\ \ &
\Omega^0_c\
\end{array}  \right)\
\label{eq:e10}
\end{eqnarray}
For spin $3/2$ sextet baryons the matrix is similar.
It is sometimes convenient to express antitriplet baryons as
\FL
\begin{eqnarray}
T_{i}=\frac{1}{2} \epsilon_{ijk}\ B_{\bar 3 jk} \ \ \ {\textstyle or}
\ \
B_{\bar 3 ij}=T_k\ \epsilon_{ijk}
\label{eq:e11}
\end{eqnarray}
Thus $T^{(c)}_1=\Xi^0_c$, $T^{(c)}_2=-\Xi^+_c$ and
$T^{(c)}_3=\Lambda^+_c$ for charmed baryons.

With respect to Dirac index, $B_{ {\bar 3}, 6}$ are $4\times4$
matrices. The spin $1/2$ and $3/2$ sextet baryon fields
with the same quark content can be combined together:
\FL
\begin{eqnarray}
B_6^\mu=\frac{\gamma_\mu+v_\mu}{\sqrt3} \gamma_5 B_6(1/2)+
B^\mu_6(3/2)
\label{eq:e12}
\end{eqnarray}
A Dirac index $\mu$ for sextet baryons is needed to
indicate the spin of light quarks \cite{r6}.
In this letter, however, we sometimes
omit this index for simplicity.
The covariant derivative $D^\mu$ acts on baryon fields similarly
to (\ref{eq:e6}):
\FL
\begin{eqnarray}
D^\mu B_{ {\bar 3}, 6}=\partial^\mu B_{ {\bar 3}, 6}
+V^\mu B_{ {\bar 3}, 6}+B_{ {\bar 3}, 6} V^{\mu T}
\label{eq:e13}
\end{eqnarray}

Under chiral $SU(3)_L\times SU(3)_R$ transformation, the various
quantities given above transform as:
\FL
\begin{eqnarray}
&&H \rightarrow H U^+, \ \ \ \
{\bar H} \rightarrow U {\bar H}  \nonumber\\
&&B_{ {\bar 3}, 6} \rightarrow U B_{ {\bar 3}, 6} U^T, \ \ \ \
{\bar B}_{ {\bar 3}, 6} \rightarrow U^{T +}
{\bar B}_{ {\bar 3}, 6} U^+  \nonumber\\
&&\xi \rightarrow L\xi U^+=U \xi R^+, \ \ \ \
\xi^+ \rightarrow U \xi^+ L^+=R \xi^+ U^+ \nonumber\\
&& D^\mu H \rightarrow (D_\mu H)  U^+ , \ \
D^\mu B_{ {\bar 3}, 6} \rightarrow U (D^\mu B_{ {\bar 3}, 6}) U^T,
\ \ \ A^\mu \rightarrow U A^\mu U^+
\label{eq:e14}
\end{eqnarray}
In (\ref{eq:e14}) we have used the row
matrix form for the heavy meson field: $H=(H_1, H_2, H_3)$.
The lowest-order chiral effective
Lagrangian in (\ref{eq:e1}) is invariant under the transformations
in (\ref{eq:e14}). It is also invariant under heavy quark spin
transformation \cite{r1,r2,r3}.

\begin{center}
{\bf 3. Chiral effective Hamiltonian of weak nonleptonic
decays of bottom hadrons}
\end{center}

In this section we write down the chiral effective Hamiltonian
for some typical weak nonleptonic decays of bottom hadrons.

We consider here the Cabibbo-favored decays of type $\Delta b=-1,
\Delta c=0, \Delta s=1$, i.e. decays of type
$ b \rightarrow c {\bar c} s$ in the free quark case.
The effective
Hamiltonian in which both $b$ quark and $c$ quark are considered
as heavy in comparison with the QCD scale $\Lambda_{QCD}$ has
been studied in \cite{r9}:
\FL
\begin{eqnarray}
&&H_{eff}=\frac{G_F V_{bc} V_{cs} }{\sqrt2}
( C_1^{\prime\prime} Q_1^{\prime\prime}+
C_2^{\prime\prime} Q_2^{\prime\prime}+...) \nonumber\\
&&O_1^{\prime\prime}=({\bar s} \Gamma_\mu h^{(b)}_v)
({\bar h^{(c)} }_{v^\prime} \Gamma^\mu h^{(\bar c)}_{\bar v})\ ,
\ \ \ \
O_2^{\prime\prime}=({\bar s_\beta} \Gamma_\mu h^{(b)}_{\alpha v})
({\bar h^{(c)}}_{\alpha v^\prime}
\Gamma^\mu h^{(\bar c)}_{\beta \bar v})\ \ \ \ \
\label{eq:e15}
\end{eqnarray}
with $\Gamma_\mu=\gamma_\mu (1-\gamma_5) $.
In (\ref{eq:e15}),
$\alpha$ and $\beta$ are color indices,
$C_1^{\prime\prime}$ and $C_2^{\prime\prime}$ are
Wilson coefficients,
$h^{(b)}_v$ indicates the heavy quark field for $b$ quark with
velocity $v$, and the ellipsis represents terms which
can be neglected in our discussion.

It is obvious that $H_{eff}$ in (\ref{eq:e15}) transforms under
$SU(3)_L\times SU(3)_R$ as $(1_R, 1_L)\otimes(1_R, \bar 3_L)=
(1_R, \bar 3_L)$. Knowing the transformation property of $H_{eff}$,
we can then construct the lowest-order chiral effective Hamiltonian
on the hadron level which transforms as $(1_R, \bar 3_L)$.

\begin{center}
{A. BOTTOM BARYON DECAYS}
\end{center}

We first study decays of a bottom baryon to a charmed baryon
, a D meson and Goldstone bosons. We denote this decay by
$B^{(b)} \rightarrow B^{(c)}+ D\ ($or $ D^\ast)+\xi$,
where $B^{(b,c)}$
represent bottom and charmed baryons, and $\xi$ represents one
or more Goldstone bosons. Let us consider the
Goldstone index for the moment.
According to the transformation properties given in
(\ref{eq:e14}), we find that there are two possible ways to form a
chiral effective Hamiltonian corresponding to (\ref{eq:e15}) which
transforms as $(1_R, \bar 3_L)$:
\FL
\begin{eqnarray}
&& H_{eff}\Rightarrow {\cal H}={\cal H}_1+{\cal H}_2\nonumber\\
&&{\cal H}_1 \sim Tr_G ( \bar B^{(c)} B^{(b)}) (H^{(c)} \xi^+)_i
\nonumber\\
&&{\cal H}_2 \sim ( H^{(c)} B^{(b)} {\bar B}^{(c)} \xi^+ )_i
\label{eq:e16}
\end{eqnarray}
where $i=3$ corresponds to the $s$ quark in (\ref{eq:e15}). If
one studies the Cabibbo-suppressed decays such as those
described by (\ref{eq:e15}) with the $s$ quark being replaced by
a $d$ quark, $i$ in (\ref{eq:e16}) should be equal to 2. The baryon
fields $B^{(c)}$ and $B^{(b)}$ could be either sextet or antitriplet
baryons. Note that $B_6$ and $B_{\bar 3}$ transform the same
way under $SU(3)_L\times SU(3)_R$ (See (\ref{eq:e14}) ).
(\ref{eq:e16}) is the lowest-order chiral effective Hamiltonian
in both Goldstone-momentum and $1/m_Q$.

Consider now the structure of ${\cal H}_1$ and ${\cal H}_2$ of
(\ref{eq:e16}) in different cases:\\
(1):
$B^{(b)}_{\bar 3} \rightarrow B^{(c)}_{\bar 3}+
D\ ($or $D^\ast)+\xi$
\FL
\begin{eqnarray}
{\cal H}_1  &&\sim
Tr_G ( \bar B^{(c)}_{\bar 3} B^{(b)}_{\bar 3} ) (H^{(c)} \xi^+)_3
=-2\  {\bar T^{(c)} }_i H^{(c)}_j \xi^+_{j3} \ T^{(b)}_i \nonumber\\
{\cal H}_2  &&\sim
(H^{(c)} B^{(b)}_{\bar 3} \bar B^{(c)}_{\bar 3} \xi^+ )_3
= {\bar T^{(c)} }_j H^{(c)}_j \xi^+_{i3} \ T^{(b)}_i
 -{\bar T^{(c)} }_i H^{(c)}_j \xi^+_{j3} \ T^{(b)}_i
\label{eq:e17}
\end{eqnarray}
It is easy to see that the term ${\cal H}_1$ corresponds to
the contribution calculated using the factorization assumption.\\
(2):
$B^{(b)}_{\bar 3} \rightarrow B^{(c)}_{6}+ D\ ($or $ D^\ast)+\xi$
\FL
\begin{eqnarray}
{\cal H}_1 &&\sim
Tr_G ( \bar B^{(c)}_{6} B^{(b)}_{\bar 3} ) (H^{(c)} \xi^+)_3
=0\nonumber\\
{\cal H}_2  &&\sim
(H^{(c)} B^{(b)}_{\bar 3}\  \bar B^{(c)}_{6}\  \xi^+ )_3
=\epsilon_{ijk}\
{\bar B^{(c)}}_{6 kl}\ H^{(c)}_j \xi^+_{l3}\ T^{(b)}_i
\label{eq:e18}
\end{eqnarray}
Thus ${\cal H}_1$ does not contribute to
$B^{(b)}_{\bar 3} \rightarrow B^{(c)}_{6}+ D\ ($or $ D^\ast)+\xi$.
\\
(3):
$B^{(b)}_{6} \rightarrow B^{(c)}_{\bar 3}+ D\ ($or $ D^\ast)+\xi$
\FL
\begin{eqnarray}
{\cal H}_1 &&\sim
Tr_G ( \bar B^{(c)}_{\bar 3} B^{(b)}_{6} ) (H^{(c)} \xi^+)_3
=0\nonumber\\
{\cal H}_2  &&\sim
(H^{(c)} B^{(b)}_{6}\  \bar B^{(c)}_{\bar 3}\  \xi^+ )_3
=\epsilon_{ikl}\ {\bar T^{(c)}_i} H_j \ \xi^+_{l3} \ B^{(b)}_{6 jk}
\label{eq:e19}
\end{eqnarray}
(4):
$B^{(b)}_{6} \rightarrow B^{(c)}_{6}+ D\ ($or $ D^\ast)+\xi$
\FL
\begin{eqnarray}
{\cal H}_1 &&\sim
Tr_G ( \bar B^{(c)}_{6} B^{(b)}_{6} )\ (H^{(c)} \xi^+)_3
={\bar B}^{(c)}_{6 ji}\ H^{(c)}_k \xi^+_{k3}\  B^{(b)}_{6 ij}
\nonumber\\
{\cal H}_2  &&\sim
(H^{(c)}\  B^{(b)}_{6}\  \bar B^{(c)}_{6}\  \xi^+ )_3
={\bar B}^{(c)}_{6 jk}\ H^{(c)}_i \xi^+_{k3}\  B^{(b)}_{6 ij}
\label{eq:e20}
\end{eqnarray}

Many $SU(3)$ relations for two-body nonleptonic decays of
type $B^{(b)} \rightarrow B^{(c)}+ D\ ($or $ D^\ast)$ can be
obtained immediately by setting $\xi=1$. We have checked that
these relations agree with those existing in the literature
\cite{r10}.

Having established the dependence of the chiral effective
Hamiltonian on the Goldstone index, we now incorporate
the heavy quark spin symmetry
in the chiral effective Hamiltonian.
\\
(1) $B^{(b)}_{\bar 3}(v) \rightarrow
B^{(c)}_{\bar 3} (v^\prime) +
D({\bar v})\ ($or $ D^\ast({\bar v})\ )+\xi$
\\
Generally, $\Lambda_b^0 \rightarrow \Lambda_c^+ D^-_s$
has two independent amplitudes and
$\Lambda_b^0 \rightarrow \Lambda_c^+ D^{-\ast}_s$ four.
In \cite{r9}, it was shown that heavy quark spin symmetry
relates $\Lambda_b^0 \rightarrow \Lambda_c^+ D^-_s$ to
$\Lambda_b^0 \rightarrow \Lambda_c^+ D^{-\ast}_s$. From \cite{r9}
the most general form of the decay amplitude for
$\Lambda_b^0 \rightarrow \Lambda_c^+ D^-_s$
consistent with heavy spin symmetry is
\FL
\begin{eqnarray}
\langle \Lambda^+_c\ D^-_s \mid H_{eff}\mid \Lambda^0_b\rangle
={\bar u_{\Lambda_c} }\Gamma_\mu
\gamma_5 \frac{1+{\not\!\bar v}}{2} N\ \Gamma^\mu\ u_{\Lambda_b}
\label{eq:e21}
\end{eqnarray}
where $\Gamma_\mu=\gamma_\mu (1-\gamma_5)$ and
$N$ is the most general $4\times 4$ matrix that can be
constructed from $v, v^\prime, {\bar v}$ and $\gamma$-matrices.
Following the arguments given in \cite{r9} to eliminate
$v^\prime$ and ${\bar v}$, one writes
\FL
\begin{eqnarray}
N=A+B {\not\! v}
\label{eq:e22}
\end{eqnarray}
Similarly, for the amplitude for
$\Lambda_b^0 \rightarrow \Lambda_c^+ D^{-\ast}_s$,
\FL
\begin{eqnarray}
\langle \Lambda^+_c\ D^{-\ast}_s \mid H_{eff}\mid \Lambda^0_b\rangle
={\bar u_{\Lambda_c} } \Gamma_\mu
{\not\!\!\epsilon^\ast} \frac{1+{\not\!\bar v}}{2}
\ N\ \Gamma^\mu u_{\Lambda_b}
\label{eq:e23}
\end{eqnarray}

The amplitude for decays involving a $\xi$-field in the final state
can be described by an effective Hamiltonian
\FL
\begin{eqnarray}
{\cal H} &&={\cal H}_1+{\cal H}_2+... \nonumber\\
{\cal H}_1  &&
={\bar T^{(c)} }_i\ \Gamma_\mu\  H^{(c)}_j\  N_1\
\Gamma^\mu \ T^{(b)}_i \xi^+_{j3} \nonumber\\
{\cal H}_2  &&=
{\bar T^{(c)} }_j \ \Gamma_\mu\
H^{(c)}_j \ N_2 \ \Gamma^\mu\  T^{(b)}_i \xi^+_{i3}
\label{eq:e24}
\end{eqnarray}
The ellipsis represents terms involving derivatives of $\xi$.
Note also that we have dropped the second term in ${\cal H}_2$
of (\ref{eq:e17})
since it is the same as ${\cal H}_1$. $N_{1,2}$ in (\ref{eq:e24})
have the same structure as (\ref{eq:e22}):
\FL
\begin{eqnarray}
N_{1,2}=A_{1,2}+B_{1,2} {\not\! v}
\label{eq:e25}
\end{eqnarray}

A specific flavor combination in decays
$B^{(b)}_{\bar 3} \rightarrow B^{(c)}_{\bar 3}+
D\ ($or $ D^\ast) +\xi$
can receive contribution from either ${\cal H}_1$ or
${\cal H}_2$ or both. If both ${\cal H}_1$ and ${\cal H}_2$
contribute, one can combine $N_1$ and $N_2$ into
$N_1+N_2=A+B {\not\!v}$, where $A_1+A_2=A$ and $B_1+B_2=B$.
\\
(2)
$B^{(b)}_{\bar 3}(v) \rightarrow B^{(c)}_{6} (v^\prime) +
D({\bar v})\ ($or $ D^\ast({\bar v})\ )+\xi$
\\
The heavy quark spin symmetry constraints on these decays
can similarly be built in by using the effective Hamiltonian
\FL
\begin{eqnarray}
{\cal H}
=\epsilon_{ijk}\ {\bar B^{(c)\rho}}_{6 kl}\ \Gamma_\mu\ H^{(c)}_j\
N_\rho\ \Gamma^\mu\ T^{(b)}_i \xi^+_{l3}+...
\label{eq:e26}
\end{eqnarray}
where $N_\rho$ can be most generally expressed as
\FL
\begin{eqnarray}
N_\rho=D_1 v_\rho+D_2 v_\rho {\not\!v}+D_3 \gamma_\rho
+D_4 \gamma_\rho {\not\!v}
\label{eq:e27}
\end{eqnarray}
Thus all decays in this class can be determined, to the lowest order,
by 4 independent amplitudes.
\\
(3)
$B^{(b)}_{6}(v) \rightarrow B^{(c)}_{\bar 3} (v^\prime) +
D({\bar v})\ ($or $ D^\ast({\bar v})\ )+\xi$
\\
Analogous to (\ref{eq:e26}) these decays are generated in the
lowest order in Goldstone boson momenta by
\FL
\begin{eqnarray}
{\cal H}
=\epsilon_{ijk} {\bar T^{(c)}_i}\ \Gamma_\mu\
H_j\ N^\prime_\rho\ \Gamma^\mu\ B^{(b)\rho}_{6 jk}\ \xi^+_{l3}+...
\label{eq:e28}
\end{eqnarray}
where
\FL
\begin{eqnarray}
N^\prime_\rho=E_1 v^\prime_\rho+E_2 v^\prime_\rho {\not\!v}+
E_3 \gamma_\rho+E_4 \gamma_\rho {\not\!v}
\label{eq:e29}
\end{eqnarray}
(4)
$B^{(b)}_{6}(v) \rightarrow B^{(c)}_{6} (v^\prime) +
D({\bar v})\ ($or $ D^\ast({\bar v})\ )+\xi$
\FL
\begin{eqnarray}
&&{\cal H}={\cal H}_1+{\cal H}_2+... \nonumber\\
&&{\cal H}_1={\bar B}^{(c)\rho}_{6 ji}\ \Gamma_\mu\
H^{(c)}_k\ N_{\rho\lambda}\ \Gamma^\mu
\ B^{(b)\ \lambda}_{6 ij}\ \xi^+_{k3}
\nonumber\\
&&{\cal H}_2={\bar B}^{(c)\ \rho}_{6 jk}\ \Gamma_\mu\ H^{(c)}_i
\ N^\prime_{\rho\lambda}\
\Gamma^\mu\ \ B^{(b)\ \lambda}_{6 ij}\ \xi^+_{k3}
\label{eq:e30}
\end{eqnarray}
where $N^{\rho\lambda}$ can be expressed as
\FL
\begin{eqnarray}
N_{\rho\lambda}=&&
g_{\rho\lambda} (F_1+F_2 {\not\!v})+
\sigma^{\rho\lambda}(F_3+F_4 {\not\!v}) \nonumber\\
&&+{\bar v}_\rho {\bar v}_\lambda (F_5+F_6 {\not\!v})+
\gamma_\rho {\bar v}_\lambda (F_7+F_8 {\not\!v})+
\gamma_\lambda {\bar v}_\rho (F_9+F_{10} {\not\!v})
\label{eq:e32}
\end{eqnarray}
$N^\prime_{\rho\lambda}$ has the same structure.

To get explicit expressions for the amplitudes with
Goldstone bosons in the final state, one uses the expansion
\FL
\begin{eqnarray}
\xi^+=1-\frac{i}{f} M-\frac{1}{f^2} M^2+...
\label{eq:e33}
\end{eqnarray}
If $\xi^+=1$ is put in (\ref{eq:e24}) to (\ref{eq:e30}),
one gets the most general effective Hamiltonian for
$B^{(b)} \rightarrow B^{(c)}+ D\ ($or $ D^\ast)$
decays constrained by
$SU(3)$ symmetry and heavy quark spin symmetry.

If $\xi^+=(-\frac{\textstyle i}{\textstyle f}) M$ is substituted into
(\ref{eq:e24}) to (\ref{eq:e30}), one obtains the lowest-order
contribution for decays involving one Goldstone boson, the
so-called ``contact term" (or commutator term in the language
of current algebra). For example, the contact term for
$\Lambda_b^0 \rightarrow \Lambda_c^+ {\bar D}^0 K^-$ is:
\FL
\begin{eqnarray}
\langle \Lambda^+_c\ {\bar D}^0 \ K^-\mid {\cal H} \mid
\Lambda^0_b\rangle_{\textstyle contact}
=(-\frac{i}{f})
\langle \Lambda^+_c\ {\bar D}^0 \mid {\cal H} \mid
\Lambda^0_b\rangle
\label{eq:e34}
\end{eqnarray}

The next-order contribution comes from the derivative terms we
have neglected and the pole terms.
In subsection D
we will give an example of how to calculate the pole terms.

\begin{center}
B. $B$-MESON DECAYS TO CHARMED BARYON ANTICHARMED BARYON PAIR AND
GOLDSTONE BOSONS
\end{center}

The structure of chiral effective Hamiltonian for $B$-meson
decays to charmed baryon-anticharmed baryon
pair and Goldstone bosons
$B \rightarrow B^{(c)} {\bar B^{(c)} } \xi$ are very similar
to those discussed above. One simply needs to make
the following substitutions:
\FL
\begin{eqnarray}
B^{(b)}_{\bar 3}\rightarrow B^{(c)}_{\bar 3},\ \ \ \ \ \
B^{(b)}_6\rightarrow B^{(c)}_6,\ \ \ \ \ \
H^{(c)} \rightarrow H^{(b)} \nonumber
%\label{eq:e34}
\end{eqnarray}
Also, the matrices
$N_1, N_2, N_\rho, N^\prime_\rho, N_{\rho\lambda}$ and
$N^\prime_{\rho\lambda}$ have similar structures. The constraints
from the heavy spin symmetry on
$B$-meson decays to charmed baryon-anticharmed baryon pairs
$B \rightarrow B^{(c)} {\bar B^{(c)} }$ were
first studied in \cite{r12}.

\begin{center}
C. $B$-MESON DECAYS TO TWO CHARMED MESONS AND GOLDSTONE BOSONS
\end{center}

Here we discuss $B$-meson decays
$B \rightarrow  D^{(\ast)} D^{(\ast)}_s  \xi$
($D^{(\ast)}=D$ or $D^\ast$) produced by
(\ref{eq:e15}). Requiring again that the chiral effective
Hamiltonian transforms as $(1_R, \bar 3_L)$ under
$SU(3)_L\times SU(3)_R$, we find (paying attention only
to Goldstone indices):
\FL
\begin{eqnarray}
{\cal H}    &&={\cal H}_1+{\cal H}_2+...\nonumber\\
{\cal H}_1  &&\sim
( H^{(b)} {\bar H}^{(c)} ) (H^{(c)} \xi^+)_3\nonumber\\
{\cal H}_2  &&\sim
( H^{(c)} {\bar H}^{(c)} ) (H^{(b)} \xi^+)_3
\label{eq:e35}
\end{eqnarray}
There are five flavor combinations for two-body decays, obtained
by setting $\xi=1$ in (\ref{eq:e35}):
$B^- \rightarrow  D^{0} D^{-}_s$ and
${\bar B}^0 \rightarrow  D^{+} D^{-}_s$ receive contribution from
${\cal H}_1$ only, ${\bar B}^0_s \rightarrow  D^{+} D^{-}$ and
${\bar B}^0_s \rightarrow  D^{0} {\bar D}^{0}$
from ${\cal H}_2$ only, and
${\bar B}^0_s \rightarrow  D^{+}_s D^{-}_s$ from both
${\cal H}_1$ and ${\cal H}_2$. The symmetry relations among
the amplitudes of these decays are the same as those
given in \cite{r11}. These symmetry relations are also true for
decays involving vector $D$-mesons. Note
that ${\cal H}_1$ corresponds to the contribution from
factorization.

Heavy quark spin symmetry can relate the amplitudes of
bottom baryon decays discussed in previous subsections.
For example, there are two independent decay amplitudes for
$\Lambda_b^0 \rightarrow \Lambda_c^+ D^-_s$ and
$\Lambda_b^0 \rightarrow \Lambda_c^+ D^{-\ast}_s$.
In \cite{r13}, it was found that there is no relation
among decay amplitudes of decays
$B \rightarrow  D D_s$, $B \rightarrow  D D^\ast_s$,
$B \rightarrow  D^\ast D_s$ and
$B \rightarrow  D^\ast D^\ast_s$.
%(However, $\Gamma(B \rightarrow  D D_s)=
%\Gamma(B^\ast \rightarrow  D D^\ast_s$)).
Thus, it is not useful
to write the chiral effective Hamiltonian for
$B \rightarrow  D^{(\ast)} D^{(\ast)}_s \xi$ in the compact form
as we did for bottom baryon decays, though this is possible.

We now look at $B$-meson decays to two charmed mesons plus one
Goldstone boson. The lowest-order contribution to such
decays comes from the contact term, the dependence of which
on Goldstone index is obtained by setting
$\xi^+=(-\frac{\textstyle i}{\textstyle f})\ M$
in (\ref{eq:e35}). For example, for
$B^- \rightarrow  D^0 {\bar D}^0 K^-$, the contact term is
\FL
\begin{eqnarray}
\langle D^0 {\bar D}^0 K^-\mid {\cal H} \mid B^-
\rangle_{\textstyle contact}=(-\frac{i}{f})
\langle D^0 {\bar D}^0 \mid {\cal H} \mid B^-
\rangle
\label{eq:e37}
\end{eqnarray}
It is easy to see from (\ref{eq:e35}) that decays with
one pion, $B \rightarrow  D^{(\ast)} D^{(\ast)}_s \pi$,
do not have the contact term.

\begin{center}
D. POLE TERMS
\end{center}

We now discuss the pole terms. Here,
as examples, we give the expressions of pole contributions,
for two decays :
$B^- \rightarrow  D^{+} D^{-}_s \pi^-$ and
$B^- \rightarrow  D^{+\ast} D^{-}_s \pi^-$. The pole diagrams
for these two decays are shown in Fig. 1. Note the $D^0$-pole
in Fig. 1b does not contribute to
$B^- \rightarrow  D^{+} D^{-}_s \pi^-$.
We first define
\FL
\begin{eqnarray}
&&\langle D^+ D^-_s \mid {\cal H} \mid {\bar B}^{0\ast} \rangle
=e^B_\mu {\cal M}^\mu_{D^+ D^-_s} \nonumber\\
&&\langle D^{+\ast} D^-_s \mid {\cal H} \mid {\bar B}^{0\ast}\rangle
=e^B_\mu \epsilon^\ast_{D^{+\ast} \nu}{\cal M}^{\mu\nu}_{D^{+\ast}
D^-_s}
\nonumber\\
&&\langle D^{0\ast} D^-_s \mid {\cal H} \mid B^{-} \rangle
=e^D_\mu {\cal N}^\mu_{D^{0\ast} D^-_s} \nonumber\\
&&\langle D^{0} D^-_s \mid {\cal H} \mid B^{-} \rangle
= {\cal N}_{D^{0} D^-_s}
\label{eq:e38}
\end{eqnarray}
where $e^B_\mu$, $e^D_\mu$ and $\epsilon^\ast_{D^{+\ast} \nu}$
are polarization vectors of
${\bar B}^{0\ast}$, $D^{0\ast}$ and $D^{+\ast}$, the vector
pole mesons
in Fig. 1. As mentioned before, heavy quark spin symmetry
does not relate the amplitudes in (\ref{eq:e38}). We find, for
$B^- \rightarrow  D^{+} D^{-}_s \pi^-$,
\FL
\begin{eqnarray}
&&\langle D^+ D^-_s \pi^-\mid {\cal H} \mid B^-
\rangle_{ {\bar B}^{0\ast}-{\textstyle pole} }
=i(\frac{g_1}{f} )
\frac{(p_\pi\cdot v)v_\mu-p_{\pi\mu} }{v\cdot p_\pi+\Delta_b}
{\cal M}^\mu_{D^+ D^-_s} \nonumber\\
&&\langle D^+ D^-_s \pi^-\mid {\cal H} \mid B^-
\rangle_{ D^{0\ast}-{\textstyle pole} }
=-i (\frac{g_1}{f} )
\frac{(p_\pi\cdot v^\prime)v^\prime_\mu-p_{\pi\mu} }
{v^\prime\cdot p_\pi - \Delta_c}
{\cal N}^\mu_{D^{0\ast} D^-_s}
\label{eq:e39}
\end{eqnarray}
where $\Delta_b=m_{B^\ast}-m_B$ and $\Delta_c=m_{D^\ast}-m_D$,
and for $B^- \rightarrow  D^{+ \ast} D^{-}_s \pi^-$,
\FL
\begin{eqnarray}
&&\langle D^{+\ast} D^-_s \pi^-\mid {\cal H} \mid B^-
\rangle_{ {\bar B}^{0\ast}-{\textstyle pole} }
=i ( \frac{g_1}{f} )
\frac{(p_\pi\cdot v)v_\mu-p_{\pi\mu} }{v\cdot p_\pi+\Delta_b}
\epsilon^\ast_{D^{+\ast} \nu}{\cal M}^{\mu\nu}_{D^{+\ast} D^-_s}
\nonumber\\
&&\langle D^{+\ast} D^-_s \pi^-\mid {\cal H} \mid B^-
\rangle_{ D^{0\ast}-{\textstyle pole} }
=( \frac{g_1}{f} )
\frac{
\epsilon_{\mu\nu\alpha\beta}\ p^\mu_\pi
\ \epsilon^{\ast \nu}_{D^{+\ast}} v^{\prime \alpha}
   }{v^\prime \cdot p_\pi} {\cal N}^\beta_{ D^{0\ast} D^-_s }
\nonumber\\
&&\langle D^{+\ast} D^-_s \pi^-\mid {\cal H} \mid B^-
\rangle_{ D^{0}-{\textstyle pole} }
=-i ( \frac{g_1}{f} )
\frac{p_\pi\cdot\epsilon^{\ast}_{ D^{+\ast} }
   }{v^\prime \cdot p_\pi+\Delta_c}  {\cal N}_{D^0 D^-_s}
\label{eq:e40}
\end{eqnarray}
Though we use (\ref{eq:e1}) for the coupling constants, we keep
$\Delta_b \neq 0$ and $\Delta_c \neq 0$ in
(\ref{eq:e39}) and (\ref{eq:e40}). One can write down
pole contributions to
$B^- \rightarrow  D^{+     } D^{-\ast}_s \pi^-$ and
$B^- \rightarrow  D^{+ \ast} D^{-\ast}_s \pi^-$ similarly.

As inputs in (\ref{eq:e38})-(\ref{eq:e40}) one could use
$B \rightarrow  D^{(\ast)} D^{(\ast)}_s$
decay amplitudes calculated in a
factorization approximation which is in good agreement with
present data \cite{r14}.

\begin{center}
{\bf 4. Discussion}
\end{center}

In the previous section we have given
the lowest-order chiral effective
Hamiltonian for various nonleptonic decays of bottom baryons
and mesons involving Goldstone bosons. This generates
the leading term in Goldstone-momentum expansion. With this
Hamiltonian, one can write down the lowest-order contributions,
the contact term and pole terms. Obviously all such discussions
and results can only be meaningful when the Goldstone bosons
involved have small momenta. With three (or more)-body decays,
there is always a small region on Dalitz-plot where momenta
of Goldstone bosons are small. Taking
$B \rightarrow  D D_s \pi$ as an example, if one defines,
\FL
\begin{eqnarray}
&&p_{12}=p_D+p_{D_s} \ \ \
(\ (m_D+m_{D_s})^2\leq p_{12}^2\leq (m_B-m_\pi)^2\simeq m_B^2 \ )
\nonumber\\
&&p_{13}=p_D+p_\pi \ \ \
( \ (m_D+m_\pi)^2\simeq m_D^2 \leq p_{13}^2\leq
(m_B-m_{D_s})^2\  )
\nonumber\\
&&p_{23}=p_{D_s}+p_\pi \ \ \
( \ (m_{D_s}+m_\pi)^2\simeq  m_D^2 \leq p_{23}^2\leq
(m_B-m_D)^2\  )
\label{eq:e42}
\end{eqnarray}
then the region where the pion is soft corresponds to
\FL
\begin{eqnarray}
&&p_{12}^2\simeq m_B^2, \ \ \ \ p_{13}^2\simeq m^2_D,
\ \ \ \ p_{23}^2\simeq m^2_{D},
\label{eq:e41}
\end{eqnarray}
or the chiral expansion parameters
$v^\prime\cdot p_\pi/\Lambda_\chi$,
${\bar v}\cdot p_\pi/\Lambda_\chi$ and
$ v\cdot p_\pi/\Lambda_\chi$ should be small. Here
$\Lambda_\chi$ is the chiral symmetry breaking scale.
We call the above region the ``zero-recoil region". Thus, using
the lowest-order amplitudes, for example
(\ref{eq:e39}) and (\ref{eq:e40}), one can calculate the decay
distributions in the zero-recoil region.

The basic idea in this letter is to provoke applications of
chiral perturbation theory to nonleptonic decays of
heavy hadrons involving Goldstone bosons, in the zero-recoil
region. Such a region only exists in three(or more)-body
decays. The situation here is very similar to the widely
studied semileptonic decays such as $B\rightarrow \pi e \nu$,
$B\rightarrow D\pi e \nu$ and
$\Lambda_b \rightarrow \Lambda_c \pi e \nu$ \cite{r1,r2,r3,r4}.
As a start, we have given the lowest-order chiral
effective Hamiltonian for several typical processes such as
$\Lambda_b \rightarrow \Lambda_c D^{(\ast)}$,
$B\rightarrow B^{(c)} \bar B^{(c)} \xi$ and
$B \rightarrow  D^{(\ast)} D^{(\ast)}_s \xi$
($D^{(\ast)}=D$ or $D^\ast$).
Some decays mentioned here, for instance
$B \rightarrow  D^{(\ast)} D^{(\ast)}_s \pi$, may soon be measured.

Finally we want to emphasize that we have assumed the validity
of heavy quark symmetry and chiral symmetry.
It is certainly necessary, as the second
step, to investigate the symmetry-breaking effects.

\ \ \ \

This research was supported by a grant to A. N. Kamal from the
Natural Sciences and Engineering Research Council of Canada.

\figure{ Pole diagrams for decays
$B^- \rightarrow  D^{+} D^{-}_s\pi^-$ and
$B^- \rightarrow  D^{+\ast} D^{-}_s\pi^-$.
The shaded squares denote
the strong vertices while the crosses denote the weak vertices.
\label{f1}}

\end{document}